\documentclass[12pt]{article}
\usepackage[utf8]{inputenc}
\usepackage{amsmath}
\usepackage{amssymb}
\usepackage{amsthm}
\usepackage{fullpage}
\usepackage{setspace}
\usepackage{enumerate}
\usepackage{subcaption}
\usepackage{apacite}
\usepackage{graphicx}
\usepackage{relsize}
 \usepackage[para]{footmisc}
\usepackage{titling}
\usepackage{blindtext}
\usepackage{tikz}
\usepackage{multirow}

\providecommand{\keywords}[1]{\textit{Keywords:} #1}

\newcommand{\bs}[1]{\boldsymbol #1}

\newcommand{\Var}[1]{\operatorname{Var}\left[#1\right]}
\newcommand{\E}[1]{\operatorname{E}\left[#1\right]}
\newcommand{\Cov}[1]{\operatorname{Cov}\left[#1\right]}

\title{Exact adaptive confidence intervals for small areas}
\author{Kyle Burris and Peter Hoff}
\begin{document}
    \maketitle
    \begin{abstract}
    In the analysis of survey data it is of interest to estimate and quantify uncertainty about means or totals for each of several non-overlapping subpopulations, or areas. When the sample size for a given area is small, standard confidence intervals based on data only from that area can be unacceptably wide. In order to reduce interval width, practitioners often utilize multilevel models in order to borrow information across areas, resulting in intervals centered around shrinkage estimators. However, such intervals only have the nominal coverage rate on average across areas under the assumed model for across-area 
heterogeneity. The coverage rate for a given area depends on the actual value of the area mean, and can be nearly zero for areas with means that are far from the across-group average. As such, the use of uncertainty intervals centered around shrinkage estimators are inappropriate when area-specific coverage rates are desired. In this article, we propose an alternative confidence interval procedure for area means and totals under normally distributed sampling errors. This procedure not only has constant $1-\alpha$ frequentist coverage for all values of the target quantity, but also uses auxiliary information to borrow information across areas.  Because of this, the corresponding intervals have shorter expected lengths than standard confidence intervals centered on the unbiased direct estimator. Importantly, the coverage of the procedure does not depend on the assumed model for across-area heterogeneity. Rather, improvements to the model for across-area heterogeneity result in reduced expected interval width.  
\end{abstract}


 \section{Introduction}
Studies that gather data from non-overlapping areas (subpopulations) are common in a variety of disciplines, including ecology \cite{brewer07}, education \cite{wall04}, epidemiology \cite{ghosh99}, and public policy \cite{maples17}.  As policy interventions have become more targeted, the demand for precise estimates of population characteristics of these areas has increased. To estimate target quantities, sample surveys may use ``direct" estimators, which are only based on the area-specific sample data. Direct estimators typically utilize survey weights, with corresponding inferences made based on the sampling design \cite{rao15}. When the direct estimates are area-specific sample averages (possibly weighted),  
the central limit theorem justifies the area-specific \textit{sampling model} 
$y_j \sim N(\theta_j, \sigma^2_j),\  j = 1, \hdots m$,
where $y_j$ is a design-unbiased and consistent direct estimate of $\theta_j$, the $j$th area mean, and $\sigma^2_j$ is the variance of the direct estimate under the sampling design. If additionally the survey data are sampled independently across areas, the joint sampling model for the area-specific direct estimates is 
\begin{equation} \label{samplingmodel}
 \bs{y} \sim N(\bs{\theta}, \bs{D}),
\end{equation}
where $\bs{y} = (y_1, \ldots, y_m)$, $\bs{\theta} = (\theta_1, \ldots, \theta_m)$, and $\bs{D}$ a diagonal matrix with elements $\{\sigma_1^2, \ldots, \sigma_m^2\}$.

For a specific area $j$, when $\sigma_j^2$ is assumed known, the classical ``direct" $1-\alpha$ confidence interval for $\theta_j$ is
\begin{equation} \label{equi-tailedz}
C^j_D(\bs{y}) = \left \{\theta : y_j + \sigma_j z_{\alpha/2} < \theta <  y_j + \sigma_j z_{1 - \alpha/2} \right \},
\end{equation}
where $z_p$ is the $p$th quantile of the standard normal distribution. This direct confidence interval has the important property of \textit{area-specific coverage} under the sampling model (\ref{samplingmodel}), since

\begin{equation} \label{constcoverage}
 \Pr( \theta_j \in C^j_D(\bs{y}) \mid \boldsymbol \theta) = 1-\alpha , 
\end{equation}
for all $\boldsymbol \theta$ and $j\in 1,\ldots, m$. 

However, it is sometimes the case that there are areas with small sample sizes under the survey design, resulting in unacceptably wide direct confidence intervals \cite{pfeffermann}. When additional precision is needed, model-based estimators are used to borrow information from other areas and utilize area-level auxiliary covariates.  A statistical model for across-area heterogeneity 
is referred to as a \textit{linking model} in the small area estimation literature. For example, the popular Fay-Herriot model \cite{fayherriot} posits that 
$\theta_j\sim N(\bs x_j^\top \bs \beta, \tau^2)$,
independently across areas, where $\bs x_j$ is a vector of 
observed area-specific covariates. 
If appropriate values of $\bs\psi= (\bs \beta,\tau^2)$ were known, then 
Bayes' rule could be used to obtain the conditional distribution of $\theta_j$ given $y_j$.  From this distribution, one could 
compute a  Bayesian credible interval

 \begin{equation} \label{bayesinterval}
C^j_B(\bs{y}) = \left \{\theta : \check \mu_j + \check \tau_j z_{\alpha/2} < \theta <  \check \mu_j + \check \tau_j z_{1 - \alpha/2} \right \},
\end{equation}
where $\check \mu_j$ and $\check \tau^2_j$ are the conditional 
mean and variance of $\theta_j$ given $y_j$, respectively. 

In practice, appropriate values for the linking model parameters $\bs{\psi}$ 
are unknown.  A Bayesian approach is to place a prior 
distribution on  $\bs{\psi}$, from which the joint posterior distribution 
of $\theta_1,\ldots, \theta_m$ may be obtained \cite{youchapman06}.  
A more common approach is an empirical Bayes strategy, whereby 
``plug-in'' estimates of $\bs\psi$ are obtained from the marginal 
likelihood of $\bs \psi$, which is itself obtained by integrating 
the density of the sampling model (\ref{samplingmodel}) for $\bs y$ 
over the values of $\bs\theta$ with respect to the linking model.  
Given such an estimate $\hat{\bs\psi}$ of $\bs \psi$, 
the empirical Bayes confidence interval is given by 
\begin{equation} \label{EBInterval}
C^j_{EB}(\bs{y}) 
= \left \{ \theta:   \hat \mu_j + \hat \tau_j z_{\alpha/2} < \theta <  \hat  \mu_j + \hat \tau_j z_{1 - \alpha/2} \right \},
\end{equation}
where $ \hat \mu_j$ and $\hat\tau^2_j$ are the conditional mean 
and variance of $\theta_j$, given $y_j$ and using $\hat{\bs \psi}$ 
as the parameters in the linking model.  Adjustments are often made to $\hat\tau^2_j$ due to the uncertainty in estimating $\hat{\bs{\psi}}$.

\begin{figure}[h]
\centering
\includegraphics[width = 0.7\textwidth]{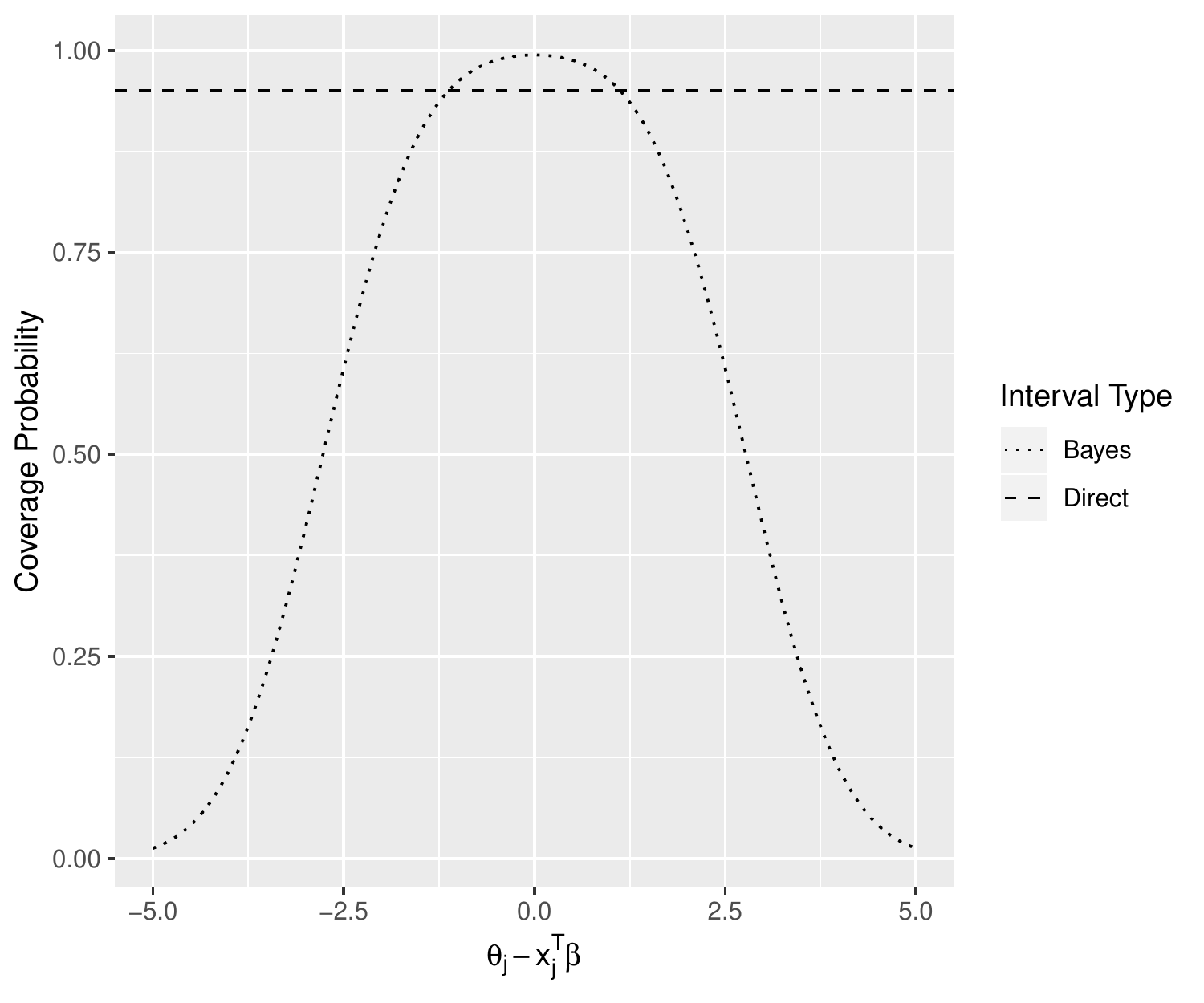}
\caption{Area-specific coverage probability for $C_D$ and $C_B$ at $\alpha = 0.05$ under the linking model $\theta_j \sim N(0, 1)$ and sampling model $y_j \sim N(\theta_j, 1)$.  Although $C_B$ (and asymptotically $C_{EB}$)  obtains 95\% coverage probability on average across values of $\theta_j - \bs x_j^\top \bs\beta$, there will be some areas that have much less than the nominal coverage probability.  In contrast, the direct interval maintains $1-\alpha$ area-specific coverage for all areas regardless of the value of $\bs\theta$. }
\label{bayes_cov}
\end{figure}

The Bayesian credible interval procedure $C_B^j$ has the property of population-level coverage, in the sense that the coverage level is $1-\alpha$ \emph{on average} with respect to the linking model.  Specifically,
\begin{equation}\label{EBgoal}
 \int \Pr( \theta_j \in C_B^j(\bs y) | \bs\theta ) \pi(\bs\theta | \bs\psi) = 1-\alpha, 
\end{equation}
where $\pi(\bs\theta|\bs\psi)$ is the probability density of $\bs\theta$ 
under the linking model. 
The empirical Bayes confidence interval procedure $C_{EB}^j$ has this property asymptotically in the number of groups, as long as $\hat{\bs{\psi}}$ is a consistent estimator of $\bs{\psi}$.  However, neither $C_B^j$ nor $C_{EB}^j$ have 
$1-\alpha$ area-specific coverage, as defined in (\ref{constcoverage}).  This is because they are centered around a biased estimator of $\theta_j$.  
To illustrate this lack of area-specific coverage, 
consider the Fay-Herriot linking model 
\label{indreglinking}
\begin{equation}
\theta_j \sim N(\bs{x}_j^\top \bs{\beta}, \tau^2),\qquad  j = 1, \ldots, m
\end{equation} 
where $\bs{x}_j$ is a vector of covariates for area $j$.  
Standard conditional probability calculations (provided in the appendix) give that 
the area-specific coverage of $C_B^j$ is a function of $\theta_j- \bs{x}_j^\top \bs{\beta}$ and can be expressed as $\Phi\left(\sigma_j (\theta_j- \bs{x}_j^\top \bs{\beta}) / \tau^2 + z_{1 - \alpha/2} \sqrt{1 + \sigma_j^2/\tau^2}\right) - \Phi\left(\sigma_j (\theta_j- \bs{x}_j^\top \bs{\beta}) / \tau^2 + z_{\alpha/2} \sqrt{1 + \sigma_j^2/\tau^2}\right)$, where $\Phi$ is the standard normal cumulative distribution function.  In general, the coverage probability for a given area will be higher than the nominal level when $\theta_j$ is close to $\bs{x}_j^\top\bs{\beta}$ and lower when $\theta_j$ is far away from $\bs{x}_j^\top\bs{\beta}$, a relationship that is visualized in Figure \ref{bayes_cov}.  This difference is amplified when the linking model variance $\tau^2$ is small relative to the sampling variance $\sigma_j^2$. 

In many applications, policy decisions and interventions are frequently targeted at outlying groups or areas.  In these cases, it is important that uncertainty intervals have area-specific coverage, so that the study has sufficient power to detect extreme values of the target quantity, regardless of what it may be.  If area-specific coverage is desired, neither the $C_B$ nor $C_{EB}$ interval procedures can be recommended, as their coverage levels will vary as a function of the target quantity $\theta_j$.  However, intervals generated by the direct interval procedure $C_D$ may also be unsatisfactory, since they may be too wide to be useful when area sample sizes are small, as they do not make use of information across areas. In this article we propose a confidence interval procedure for small area analysis that maintains exact area-specific coverage, while also allowing for information sharing across areas, thereby offering 
improved precision over direct interval procedures. Like direct confidence intervals, these intervals have exact $1-\alpha$ area-specific coverage under the sampling model (\ref{samplingmodel}), regardless of whether or not a 
particular linking model is accurate. Importantly, unlike the Bayes and empirical Bayes procedures, our procedure is appropriate for area-level inference in that it maintains area-specific coverage rates. However, 
like the Bayes and empirical Bayes intervals, our proposed intervals will be shorter than the direct intervals on average with respect to the linking model. 

Our proposed interval procedure extends that of \citeA{yuhoff16}, who developed an adaptive 
procedure with area-specific coverage using an exchangeable linking model. In this article we extend this idea to the types 
of linking models often used for small area analysis, including models that allow for area-specific features and spatial or 
temporal correlation between area means.  In Section 2, we briefly review the interval procedure first developed by \citeA{pratt63}, and extended by \citeA{yuhoff16} to include the case of unknown sampling variances.  We also demonstrate how to apply these ideas to the analysis of small areas, using the spatial Fay-Herriot linking model as a running example.  Section 3 describes a simulation study designed to compare interval procedures under a variety of linking models.  In Section 4 we apply our methodology to estimate household radon levels in 196 U.S. counties.  A discussion follows in Section 5.

\section{Methods}

\subsection{The FAB interval procedure}
We first consider constructing a $1-\alpha$ confidence interval procedure
for a specific group $j$, based on the 
sampling model (\ref{samplingmodel}), where for now we assume $\sigma_j^2$ to be known.  Let $s_j$ be any function mapping $\mathbb R$ to the 
unit interval $[0,1]$, possibly depending on 
data from other areas, that is, $\bs y_{-j} = \{y_i : i\neq j\}$. 
Then, assuming the sampling model, it is easily verified that 
\begin{equation} \label{generals_interval}
C_{s_j}^j  =  \left \{\theta : y_j + \sigma_j z_{\alpha(1 - s_j(\theta))} < \theta <  y_j + \sigma_j z_{1 - \alpha s_j(\theta)} \right \}
\end{equation}
is a valid $1-\alpha$ frequentist confidence region, satisfying the area-specific coverage property (\ref{constcoverage}).  
The standard direct interval corresponds to $s_j(\theta) =1/2$.  

Now suppose that, based on $\bs y_{-j}$ and a linking model, we believe 
$\theta_j$ is likely to be near some value $\mu_j$. We encode this 
belief with a normal 
probability distribution $\theta_j \sim N(\mu_j, \tau^2_j)$. 
For example, $\mu_j$ and $\tau^2_j$ might be the conditional expectation and 
variance of $\theta_j$, given $\bs y_{-j}$ and the linking model. Given such information, we may prefer an area-specific interval 
procedure that, relative to the direct interval,  
is more precise (has shorter expected width) for values of 
$\theta_j$ near $\mu_j$, at the expense of having longer 
expected width for values of $\theta_j$ deemed unlikely by the 
linking model. We may then wish to use the area-specific 
interval procedure that minimizes the expected width, relative 
to the linking model. 

The minimizer of this expected width among all $1-\alpha$ frequentist 
intervals can be obtained using results of 
\citeA{pratt63}, who considered frequentist interval construction 
for a single mean parameter with a normal prior distribution. 
The $1-\alpha$ frequentist interval that has minimum width, on average 
 with respect to a $N(\mu_j,\tau^2_j)$ distribution 
for $\theta_j$, can be shown to be given by 
(\ref{generals_interval}) with 
\begin{align} \label{optimal_s_fabz}
\begin{split}
s_j(\theta) &= g^{-1}(2 \sigma_j (\theta - \mu_j)/\tau_j^2)\\
g(\omega) &= \Phi^{-1}(\alpha \omega) - \Phi^{-1}(\alpha (1 - \omega)).
\end{split}
\end{align}

\begin{figure}[h]
\centering
\includegraphics[width = 0.7\textwidth]{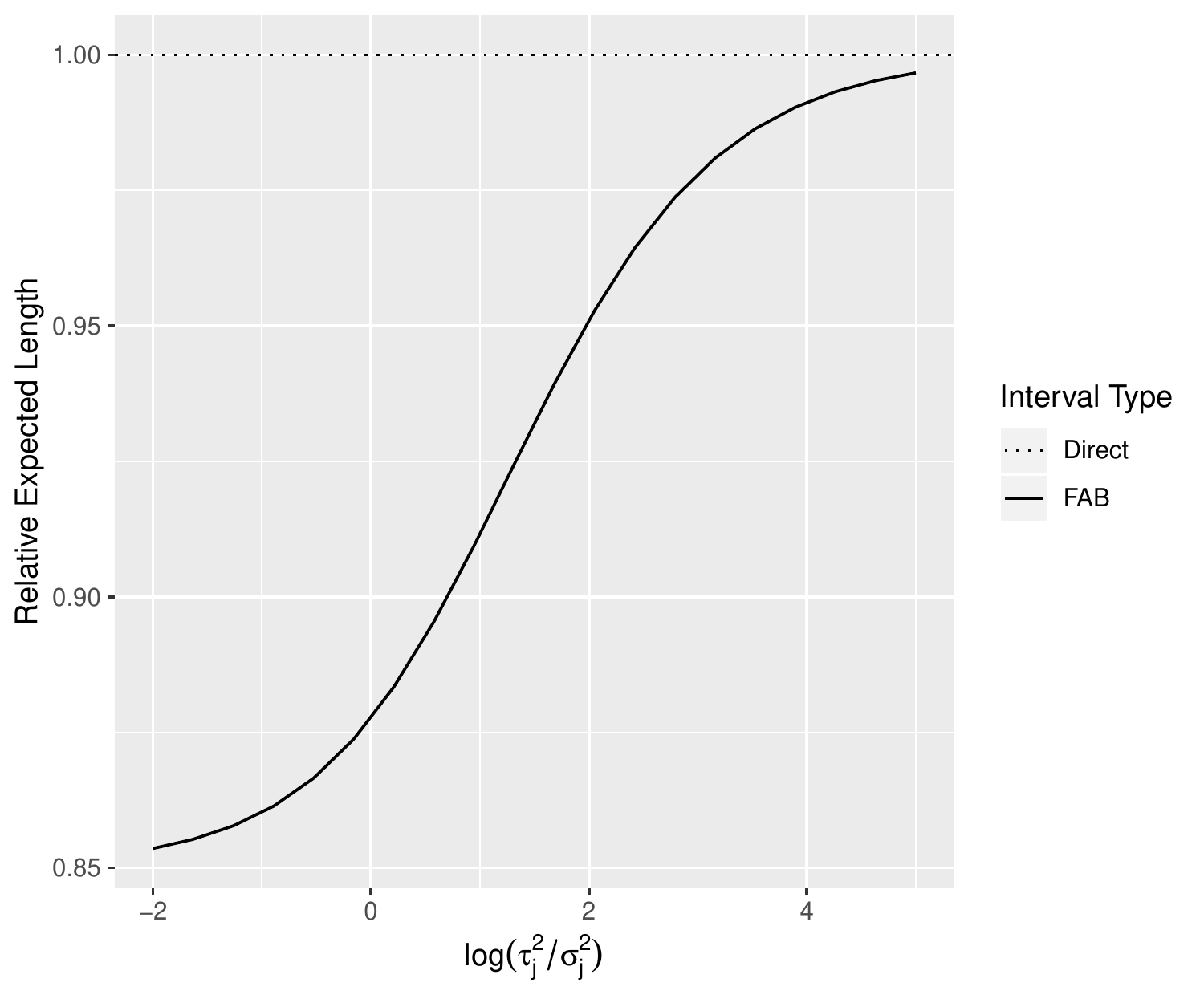}
\caption{The expected relative improvement of the 95\% FAB $z-$interval over the direct interval.  When the prior variance is of a similar magnitude to the sampling variance or smaller, there can be a substantial reduction in interval width.  However, there appear to be diminishing returns as the prior variance $\tau_j^2$ decreases, due to the constraint of constant $1-\alpha$ frequentist coverage.} 
\label{tauwidthfig}
\end{figure}

Following \citeA{yuhoff16}, we refer to confidence intervals constructed in this way as FAB intervals because,
thinking of the conditional distribution of $\theta_j$ given $\bs y_{-j}$ as a prior distribution, they are ``frequentist, assisted by Bayes''.  Importantly, even if $\theta_j$ is located in a region of low probability under the linking model, a FAB interval will still maintain $1-\alpha$ area-specific coverage for $\theta_j$.  As such, the FAB interval procedure is coverage-robust to misspecification of the linking model.   
In terms of precision, if the linking model reasonably describes the across-area heterogeneity in means, then the FAB procedure will represent an improvement over the direct procedure, on average across areas (Figure \ref{tauwidthfig}).
In contrast, the Bayes and empirical Bayes interval procedures do not maintain 
constant area-level coverage rates even if the linking model perfectly describes the across-area distribution of $\bs{\theta}$ (unless all area-specific 
means are the same).

\subsection{FAB intervals for the spatial Fay-Herriot model}
The spatial Fay-Herriot model is frequently employed by researchers and statistical agencies due to the abundance of cross-sectional survey data that come from non-overlapping geographic areas such as counties, neighborhoods, school districts, and electoral precincts.  Area-level direct estimates from this type of data typically exhibit high spatial autocorrelation, in which areas closer together tend to have similar values for their target quantities, even after accounting for the auxiliary covariates.

The spatial Fay-Herriot model includes the sampling model 
(\ref{samplingmodel}) which we assume to be correct, and 
a spatial linking model for across-unit heterogeneity of the 
$\theta_j$'s, which we do not assume is correct. The linking 
model can be written as 

\begin{equation}
\bs{\theta} = \bs{X}\bs{\beta} + \bs{u}, \qquad \bs{u} \sim N(\bs{0}, \bs{G}(\bs{\psi}))
\end{equation}
where $\bs{\psi} = \{\tau^2, \rho\}$ parameterizes the dispersion and spatial relationship of the random effects.  The conditional autoregressive (CAR) model and the simultaneous autoregressive (SAR) model are two of the main approaches for structured covariance modeling of spatially autocorrelated areal data \cite{gelfandspatialbook}.  Following \cite{singh05} and \cite{pratesisalvati08}, we consider the SAR model

\begin{equation}
\bs{u} = \rho \bs{W} \bs{u} + \bs{v} \qquad \Rightarrow \qquad \bs{u} = (\bs{I} - \rho \bs{W})^{-1}\bs{v},
\end{equation}
where $\bs{W}$ is a $m \times m$ neighborhood proximity matrix, $\rho$ a spatial relationship parameter, and $\bs{v}$ a $m \times 1$ mean-zero random vector with independent normal entries, each with variance $\tau^2$.  A binary contiguity neighborhood matrix is often chosen for $\bs{W}$, in which $W_{ij} = 1$ if areas $i$ and $j$ are neighbors and zero otherwise.  Regardless of the choice of $\bs{W}$, it is typically first row-standardized to make the row elements sum to one.  When the proximity matrix is standardized in this way, $\bs{I} - \rho \bs{W}$ is non-singular when $\rho \in (-1, 1)$, and $\rho$ can be treated as a spatial autocorrelation parameter.

Combining the above equations, the linking model for $\bs{\theta}$ becomes

\begin{equation} \label{spatiallinkingmodel}
\bs{\theta} \sim N(\bs{X}\bs{\beta}, \tau^2 [(\bs{I} - \rho \bs{W}) (\bs{I} - \rho \bs{W}^T)]^{-1})
\end{equation}

Our proposed confidence interval for a small area mean $\theta_j$ is 
obtained by first 
using data $\bs y_{-j} = (y_{1},\ldots, y_{j-1},y_{j+1},\ldots, y_m)$ from the other groups, 
along with the linking model (\ref{spatiallinkingmodel})
to obtain a mean 
$\mu_j$ and variance $\tau^2_j$ that describe the likely 
values of $\theta_j$, and then using these values to construct the 
FAB interval given by (\ref{generals_interval}) and (\ref{optimal_s_fabz}). 
Recall that
the resulting confidence interval has exact  $1-\alpha$ 
coverage for 
$\theta_j$, regardless of the value of $\theta_j$ or the 
accuracy of the linking model, 
as long as the sampling model is correct and the values of 
$\mu_j$ and $\tau^2_j$ are chosen independently of the value of 
$y_j$. 

A fully Bayesian approach to 
obtaining values of $\mu_j$ and $\tau^2_j$ 
would be to take them to be the conditional mean and variance of 
$\theta_j$ given $\bs y_j$, 
under a suitable prior distribution for the parameters $\{\bs{\beta}, \tau^2, \rho\}$ of the linking model, and computed using a Markov chain Monte Carlo 
approximation algorithm.   However, this can be prohibitively computationally costly, 
as a separate approximation would need to be run for each area.  As a more feasible alternative, we suggest an empirical Bayes approach in which 
$\{\bs{\beta}, \tau^2, \rho\}$ are first estimated from the marginal distribution of $\bs y_{-j}$, which are then used to obtain empirical Bayes estimates of the $\bs\theta_{-j}$'s. The resulting conditional mean and variance of $\theta_j$, 
using ``plug-in'' values of $\bs\theta_{-j}$ and 
  $\{\bs{\beta}, \tau^2, \rho\}$, are given by 
\begin{align} \label{conditionaltheta}
\mu_j = \E{\theta_j \mid \bs\theta_{-j} = \hat{\bs\theta}_{-j}, \bs{\beta} = \hat{\bs{\beta}}, \rho = \hat{\rho}, \tau^2 = \hat{\tau}^2} &= \bs{x}_j^\top \hat{\bs{\beta}} + \hat{\bs{G}}_{j, -j} \hat{\bs{G}}_{-j, -j}^{-1} \left(\hat{\bs{\theta}}_{-j} - \bs{X}_{-j}\hat{\bs{\beta}}\right)  \\
\tau^2_j = \Var{\theta_j \mid \theta_{-j} = \hat{\theta}_{-j}, \bs{\beta} = \hat{\bs{\beta}}, \rho = \hat{\rho}, \tau^2 = \hat{\tau}^2} &= \hat{\bs{G}}_{j, j} - \hat{\bs{G}}_{j, -j} \hat{\bs{G}}_{-j, -j}^{-1} \hat{\bs{G}}_{-j, j},  \nonumber
\end{align}
where $\hat{\bs{G}} = \hat{\tau}^2 \left(\left(\bs{I} - \hat{\rho}\bs{W}\right)\left(\bs{I} - \hat{\rho}\bs{W}\right)\right)^{-1}$ 

In sum, the steps to obtain a FAB interval are 

\begin{enumerate}
 \item Estimate linking model parameters using data from all counties 
   other than $j$. Details of maximum likelihood estimation for the spatial Fay-Herriot are provided in the Appendix.
   \item Obtain a normal prior distributions for both $\theta_j$ using plug-in estimates from the fitted linking model.  In the case of the spatial Fay-Herriot model, the prior mean $\mu_j$ and prior variance $\tau_j^2$ are given by (\ref{conditionaltheta}).
\item Obtain the optimal $s$-function for county $j$ given prior information about $\theta_j$, as described in Section 2.1.
\item Construct the FAB $z$-interval $\left \{\theta : \bar{y}_j + \sigma_j t_{\alpha(1 - s_j(\theta))} < \theta <  \bar{y}_j + \sigma_j t_{1 - \alpha s_j(\theta)} \right \}$.
\end{enumerate}

\subsection{Unknown within-area variances}
The procedure detailed above assumes that the sampling variance $\sigma^2_j$ is known (or known with a high degree of accuracy). 
However, in practice the variance of the direct estimate of each area is rarely known, and only consistent estimates $\hat{\sigma}_j^2$ are available.  Under the assumption that the response is normally distributed within area $j$,
\begin{equation} \label{samplevar}
q_j \hat{\sigma}_j^2/\sigma^2_j \sim \chi_{q_j}^2,
\end{equation}
where $q_j$ is the effective number of degrees of freedom for area $j$ implied by the sampling design \cite{Cochran77}.  \citeA{yuhoff16} extended Pratt's original $z$-interval to the case of an unknown sampling variance as follows:  If the sample statistics $y_j$ and $\hat{\sigma}_j^2$ are independent, where $y_j \sim N(\theta_j, \sigma_j^2)$ and $q_j \hat{\sigma}_j^2/\sigma_j^2 \sim \chi^2_{q_j}$, then for any nondecreasing function $s_j: \mathbb{R} \rightarrow [0,1]$,

\begin{equation} \label{general_t}
C^j_{s_j}(y, \hat{\sigma}_j) = \{\theta: y_j + \hat{\sigma}_j t_{\alpha (1 - s_j(\theta)), q_j} < \theta < y_j + \hat{\sigma}_j t_{1 - \alpha s_j(\theta), q_j}\},
\end{equation}
where $t_{p, q_j}$ is the $p$th quantile of the $t$ distribution with $q_j$ degrees of freedom, is a valid $1-\alpha$ confidence interval with area-specific coverage. The function $s_j(\theta)$ can be selected on the basis of prior information about not only the target quantity $\theta_j$, but also the sampling variance $\sigma_j^2$.  If this prior information can be summarized by a normal distribution for $\theta_j$ and an inverse-gamma distribution for $\sigma_j^2$, it is possible to obtain the function $s_j$ that minimizes the prior expected length of the interval (\ref{general_t}) via numerical methods described in \citeA{yuhoff16}.  To obtain this prior information, we recommend specifying a linking model for both $\bs{\theta}$ and $\bs{D}$, possibly allowing for the presence of auxiliary covariates in the model for $\bs{D}$.  As before, 
parameters of the linking model can be estimated and moment-matching used to
obtain a normal distribution for $\theta_j$ and an inverse-gamma distribution 
for $\sigma_j^2$, which represent the indirect information about 
$\theta_j$ with which a FAB $t$-interval may be constructed.  We provide an empirical example of the FAB $t$-interval procedure in Section 4.

\section{Simulation study}
To compare the properties of FAB intervals and direct intervals, we constructed a simulation study in which area means may exhibit spatial autocorrelation and/or association with an explanatory variable.  We aimed to quantify the reduction in expected interval width obtained via the FAB interval procedure relative to the direct interval procedure.  Throughout the study, we assumed the sampling model $y_j \sim N(\theta_j, \sigma^2_j)$ with $\sigma_j^2 = 1$ known for all areas $j$, yielding the direct confidence interval $C^j_D =  y_j \pm z_{1 - \alpha/2}$.

Forty-nine 
areas were  located on a $7 \times 7$ lattice.  For each of 5000 datasets, we simulated area means under the following procedure:
\begin{enumerate}[1)]
\item Draw $u_j \sim U(0, 1),\text{ } j = 1, \ldots, m$
\item Set $x_j = \frac{u_j - \bar{u}}{s_u}, \qquad \bar{u} = \frac{1}{m}\sum_{j=1}^m u_j, \qquad s_u = \frac{1}{m-1} \sum_{j=1}^m (u_j - \bar{u})^2 \qquad \bs{X} = (x_1, \cdots, x_m)^T$
\item Draw $\bs{\theta} \sim N\left(\bs{X}\beta, \tau^2 [(\bs{I} - \rho \bs{W}) (\bs{I} - \rho \bs{W}^T)]^{-1}\right)$
\item Draw $y_j \sim N(\theta_j, 1)$,
\end{enumerate}

This data generating procedure was repeated eight times, one for each setting of $\rho \in \{0, 0.9\}$, $\tau^2 \in \{0.5, 5\}$ and $\beta \in \{0, 10\}$.  In each repetition, the neighborhood matrix $\bs{W}$ was assumed to be a row standardized binary contiguity matrix (a binary contiguity matrix is defined such that the $i,j$th entry equals 1 if areas $i$ and $j$ border each other, and $0$ otherwise). 

\subsection{Intervals with Area-Specific Coverage}

For each area in a dataset, we constructed five types of 95\% confidence intervals that have area-specific coverage.  These consist of the direct interval and four different FAB intervals based on maximum likelihood estimation of linking models ranging in complexity.  The linking models considered were

\begin{enumerate}[1)]
\item The exchangeable model: $\theta_j \sim N(0, \tau^2)$ independently across groups $j = 1, \ldots, m$. 
\item The covariate model: $\theta_j \sim N(x_j\beta, \tau^2)$ independently across groups $j = 1, \ldots, m$. 
\item The spatial model:  $\bs{\theta} \sim N\left(\bs{0}, \tau^2 [(\bs{I} - \rho \bs{W}) (\bs{I} - \rho \bs{W}^T)]^{-1}\right)$. 
\item  The full model: $\bs{\theta} \sim N\left(\bs{X}\beta, \tau^2 [(\bs{I} - \rho \bs{W}) (\bs{I} - \rho \bs{W}^T)]^{-1}\right)$.
\end{enumerate}

Under each data generating process, average interval lengths over all simulations for these five confidence interval procedures were calculated.  Average lengths relative to the direct interval are given in Table \ref{table1} for each of the four FAB procedures.  For the simulations in which the data were generated with strong spatial autocorrelation $\rho = 0.9$, the spatial FAB intervals outperformed their non-spatial counterparts in terms of average interval width.  Similarly, the non-spatial FAB intervals are slightly narrower than their spatial counterparts when the data is generated without spatial autocorrelation due to the increased uncertainty that comes with estimating $\rho$.  For lower values of the random effect variance $\tau^2$, the FAB intervals are significantly narrower due to the increased precision of the available indirect information.  When a covariate is a strong predictor of the area mean, FAB intervals estimated under a linking model with a covariate were narrower than those without a covariate.  Most importantly, no matter the linking model, FAB intervals were narrower on average than intervals based on direct estimates alone.  The percentage decrease in interval length ranged from 0.4\% to 13.2\%.

\begin{table}[ht]
\footnotesize
\centering
\begin{tabular}{|c|c|c|c|c|c|c|c|c|}
  \hline
   & \multicolumn{4}{|c|}{$\tau^2 = 1/2$} &  \multicolumn{4}{|c|}{$\tau^2 = 5$}\\
\cline{2-9}
  & \multicolumn{2}{|c|}{$\beta = 0$} &  \multicolumn{2}{|c|}{$\beta = 10$} & \multicolumn{2}{|c|}{$\beta = 0$} &  \multicolumn{2}{|c|}{$\beta = 10$} \\
  \cline{2-9}
Linking Model & $\rho = 0$ & $\rho = 0.9$ & $\rho = 0$ & $\rho = 0.9$ & $\rho = 0$ & $\rho = 0.9$ & $\rho = 0$ & $\rho = 0.9$\\
    \hline 
    Exchangeable & 0.868 & 0.901 & 0.995 & 0.996 & 0.938 & 0.976 & 0.996 & 0.996 \\ 
         Covariate & 0.869 & 0.901 & 0.869 & 0.901 & 0.939 & 0.977 & 0.939 & 0.976 \\ 
             Spatial & 0.868 & 0.877 & 0.996 & 0.996 & 0.939 & 0.939 & 0.996 & 0.996 \\ 
                  Full & 0.869 & 0.878 & 0.869 & 0.878 & 0.940 & 0.940 & 0.940 & 0.940 \\ 
   \hline
\end{tabular}
\caption{Average confidence interval length relative to the direct interval by simulation, each with 5000 datasets.  Since each column represents a separate data generating process, interval widths should only be compared across columns.  FAB intervals are narrower on average than direct intervals and are narrower when the linking model appropriately models the data generating process.}
\label{table1}
\end{table}

It is important to note that a given FAB interval is not guaranteed to be narrower than the corresponding direct interval.  Rather, FAB intervals will be narrower on average than direct intervals.  Table \ref{table2} details the percentage of areas with shorter FAB intervals than direct intervals by simulation.  

\begin{table}[ht]
\footnotesize
\centering
\begin{tabular}{|c|c|c|c|c|c|c|c|c|}
  \hline
   & \multicolumn{4}{|c|}{$\tau^2 = 1/2$} &  \multicolumn{4}{|c|}{$\tau^2 = 5$}\\
\cline{2-9}
  & \multicolumn{2}{|c|}{$\beta = 0$} &  \multicolumn{2}{|c|}{$\beta = 10$} & \multicolumn{2}{|c|}{$\beta = 0$} &  \multicolumn{2}{|c|}{$\beta = 10$} \\
  \cline{2-9}
Linking Model & $\rho = 0$ & $\rho = 0.9$ & $\rho = 0$ & $\rho = 0.9$ & $\rho = 0$ & $\rho = 0.9$ & $\rho = 0$ & $\rho = 0.9$\\
    \hline 
    Exchangeable & 96.7\% & 91.9\% & 81.3\% & 81.5\% & 86.8\% & 83.6\% & 81.7\% & 81.9\% \\ 
         Covariate & 96.5\% & 91.4\% & 96.5\% & 91.5\% & 86.1\% & 82.7\% & 86.0\% & 82.6\% \\ 
             Spatial & 96.6\% & 95.5\% & 79.2\% & 79.4\% & 85.9\% & 88.4\% & 79.6\% & 80.2\% \\ 
                  Full & 96.4\% & 95.2\% & 96.4\% & 95.2\% & 85.1\% & 87.5\% & 85.0\% & 87.5\% \\ 
   \hline
\end{tabular}
\caption{Percentage of areas for which the FAB interval is narrower than the corresponding direct interval, by simulation.  For a vast majority of the areas, any FAB interval will be narrower than the direct interval, regardless of the linking model chosen.  However, there are more areas that demonstrate improvements when the linking model appropriately models the data generating process.}
\label{table2}
\end{table}

\subsection{Comparison to Empirical Bayes}
In addition to the direct interval and four FAB intervals, we also calculated empirical Bayes (EB) intervals based on the four linking models detailed above.  Because empirical Bayes intervals are not constrained to have area-specific coverage, they are able to be narrower than FAB intervals, particularly when  
each area-level mean is well-predicted by the linking model (e.g., 
$\tau^2$ is small). However, as shown in Table 3, 
empirical Bayes and FAB intervals have increasingly similar average widths 
as $\tau^2$ increases. 

\begin{table}[ht]
\footnotesize
\centering
\begin{tabular}{|l|c|c|c|c|c|c|c|c|c|}
  \hline
    &  & \multicolumn{4}{|c|}{$\tau^2 = 1/2$} &  \multicolumn{4}{|c|}{$\tau^2 = 5$}\\
    \cline{3-10}
   &   & \multicolumn{2}{|c|}{$\beta = 0$} &  \multicolumn{2}{|c|}{$\beta = 10$} & \multicolumn{2}{|c|}{$\beta = 0$} &  \multicolumn{2}{|c|}{$\beta = 10$} \\
        \hline
Type & Linking Model & $\rho = 0$ & $\rho = 0.9$ & $\rho = 0$ & $\rho = 0.9$ & $\rho = 0$ & $\rho = 0.9$ & $\rho = 0$ & $\rho = 0.9$\\
  \hline
  EB & Exchangeable & 2.387 & 3.182 & 3.903 & 3.903 & 3.601 & 3.816 & 3.903 & 3.905 \\ 
  EB & Covariate & 2.445 & 3.203 & 2.447 & 3.195 & 3.609 & 3.818 & 3.608 & 3.818 \\ 
  EB & Spatial  & 2.511 & 2.682 & 3.904 & 3.904 & 3.621 & 3.597 & 3.904 & 3.905 \\ 
    EB & Full & 2.576 & 2.726 & 2.580 & 2.719 & 3.633 & 3.609 & 3.632 & 3.610 \\ 
    \hline 
       FAB & Exchangeable & 3.402 & 3.530 & 3.902 & 3.902 & 3.679 & 3.826 & 3.903 & 3.905 \\ 
  FAB & Covariate & 3.405 & 3.533 & 3.405 & 3.531 & 3.682 & 3.828 & 3.682 & 3.828 \\ 
    FAB & Spatial  & 3.403 & 3.440 & 3.903 & 3.903 & 3.682 & 3.681 & 3.904 & 3.905 \\ 
        FAB & Full & 3.405 & 3.443 & 3.406 & 3.442 & 3.686 & 3.685 & 3.686 & 3.685 \\ 
   \hline
\end{tabular}
\caption{Average lengths of FAB and empirical Bayes (EB) confidence intervals by simulation.  In general, when the across-area heterogeneity is small, empirical Bayes intervals are able to be much narrower than FAB intervals.  }
\label{table3}
\end{table}

\begin{figure}[ht]
\centering
\begin{subfigure}{0.49\textwidth}
\includegraphics[width = \textwidth]{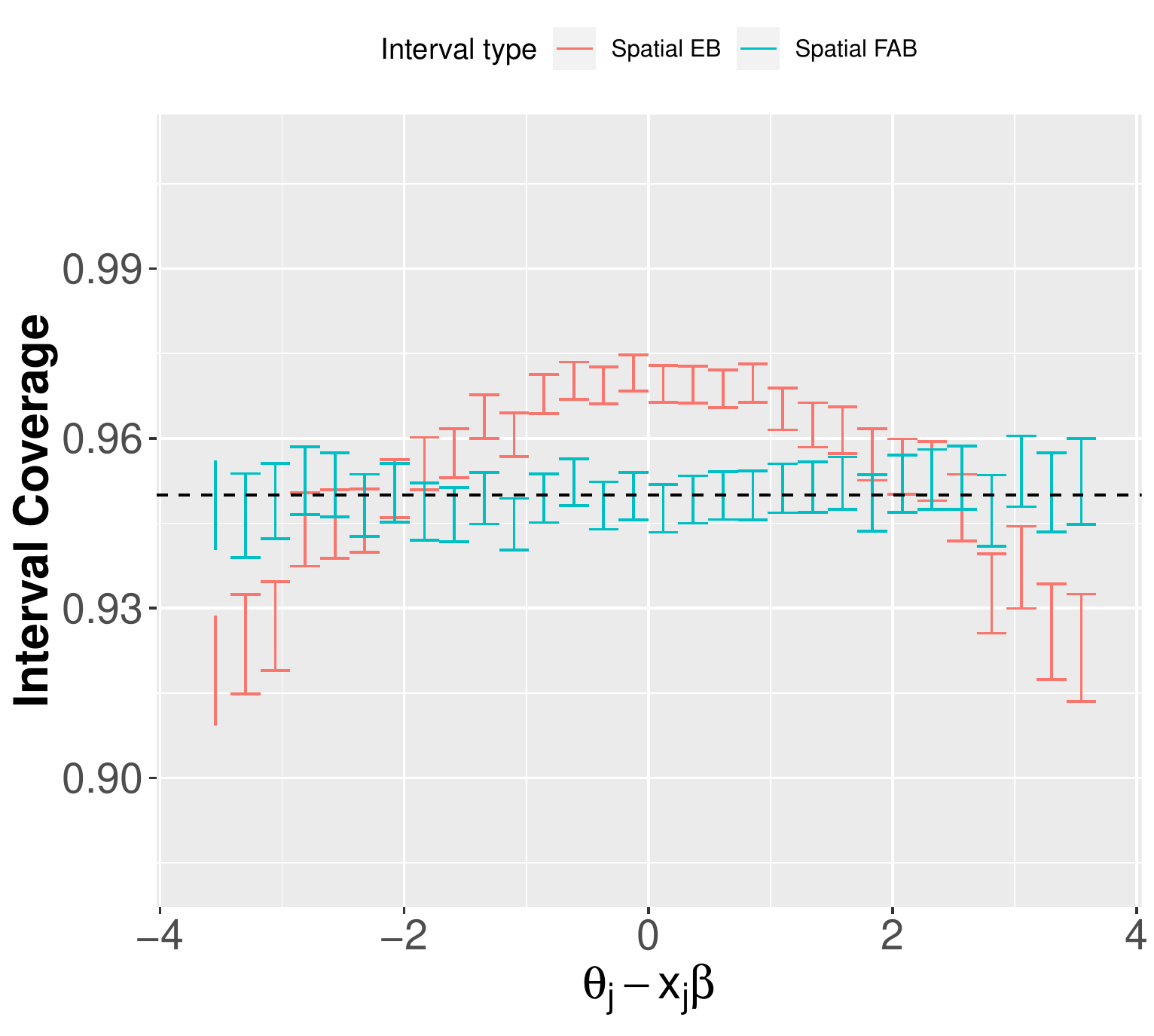}
\caption{$\rho = 0$, $\beta = 10$, $\tau^2 = 5$}
\end{subfigure}
\begin{subfigure}{0.49\textwidth}
\includegraphics[width = \textwidth]{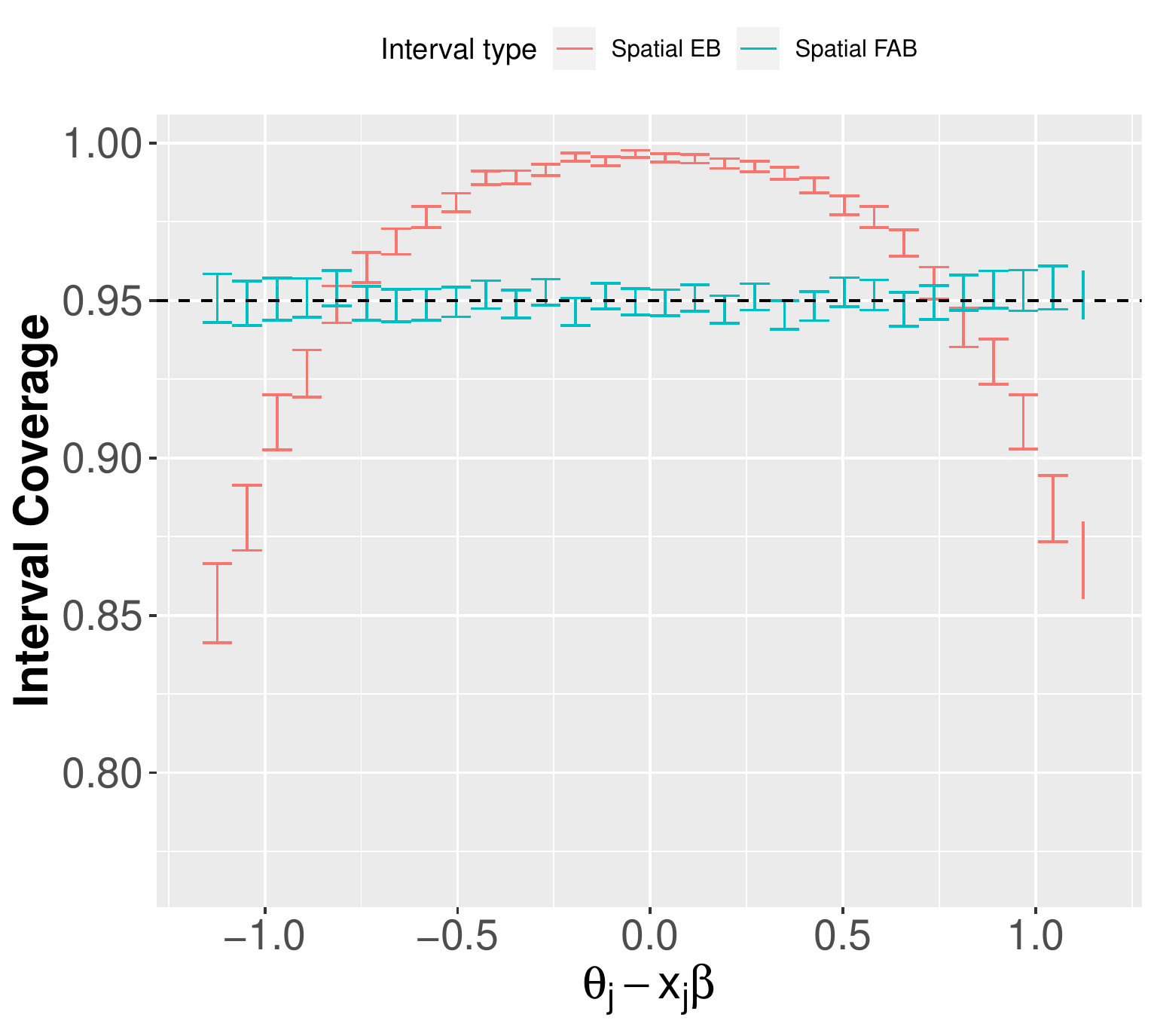}
\caption{$\rho = 0$, $\beta = 0$, $\tau^2 = 0.5$}
\end{subfigure}
\caption{Estimated coverage rate and 95\% confidence interval for binned values of $\theta_j - \bs{X}\bs{\beta}$.  It is readily apparent that there are values of $\theta_j$ for which the empirical Bayes interval has far less than $1-\alpha$ coverage.  In contrast, the FAB interval has area-specific coverage.}
\label{figure1}
\end{figure}

However, Table \ref{table3} does not tell the full story.  Although the empirical Bayes confidence intervals approximately achieve $1-\alpha$ coverage on average across areas, the actual coverage rate depends on the value of unknown target quantity $\theta_j$.  For values of $\theta_j$ that are close to their predicted means under the linking model, the EB interval has greater than $1-\alpha$ coverage.  For values much farther away, it has much less (Figure \ref{figure1}), since each EB interval is centered around a biased estimate of the target quantity.  In fact, with the exception of two points, the frequentist coverage of the EB interval is unequal to $1-\alpha$ for all values of $\theta_j$.  Unlike the EB interval, the FAB interval shares the property of constant coverage with the direct interval.  

\section{Empirical example: Household radon levels}\label{sec4}
Between 1987 and 1988, the U.S. Environmental Protection Agency collected household-level data on radon concentration as part of its State Residential Radon Survey (SRRS).  The data consist of a stratified random sample of 12,777 homes, each located in one of 472 counties in nine different states.  We examine a subset of the SRRS data, concentrating on four of the nine states in the study: Minnesota, Wisconsin, Michigan, and Indiana.  These states are geographically close and demographically similar to one another, 
so patterns of radon concentration may be similar across this region. Within these four states, there are 3,767 household measurements, located in 209 distinct counties.  One of the primary goals of the study was to ``provide the best estimate and uncertainty quantification of a county's true geometric mean of radon screening measurements'' \cite{price1996}.  

\citeA{price1996} analyzed the subset of the SRRS data from the state of Minnesota, developing a linear mixed model to construct 95\% Bayesian credible intervals for county-specific geometric mean radon levels.  However, these intervals do not have 95\% coverage at the county level.  In particular, as outlined in Section 1, they will suffer from undercoverage for counties with exceptionally high or low true means.  Such systematic undercoverage can be dangerous, because counties with extremely high radon levels present significant public health risks to their communities and need to be detected to necessitate appropriate policy action.  As such, we should use interval procedures that maintain known, constant county-specific coverage rates. 

Of the 209 counties in the data, 124 of them have fewer than ten sampled households, and 72 have fewer than five sampled households.  For these counties, direct confidence intervals will be extremely wide, which can limit their usefulness in practice.  However, the average precision of these intervals can be improved by using FAB intervals to borrow information across counties.  In this section, we compare direct intervals to several FAB interval procedures corresponding to different linking models for the county-specific means.   

Following \citeA{price1996}, we make a small empirical adjustment to the radon concentration values to mitigate the impact of very low concentration measurements that arise as a byproduct of measurement error.  In addition, we also follow the authors in assuming that adjusted radon concentrations within counties follow a roughly log-normal distribution, which appears warranted by exploratory data analysis.  Letting $y_{i,j}$ be the log adjusted radon concentration measurement for household $i$ in county $j$, we assume the within-county sampling model $y_{1,j}, \hdots, y_{n_j, j} \sim N(\theta_j, \omega_j^2)$, where $\theta_j$ is the unknown true geometric mean radon concentration for county $j$ and $\omega^2_j$ is the unknown variance of log radon measurements in county $j$.  Under the assumption of random sampling within counties, the county sample mean is distributed as $\bar{y}_j  \sim N(\theta_j, \sigma^2_j)$, where $\sigma_j^2 = \omega_j^2 / n_j$ is the variance of the sample mean.  We define $\hat{\sigma}^2_j = \hat{\omega}_j^2 / n_j$, where $\hat{\omega}_j^2$ is the sample standard deviation of log-radon measurements within county $j$. $\hat{\sigma}^2_j$ is an unbiased and consistent estimate of $\sigma_j^2$.

We illustrate the use of FAB intervals for this small area analysis 
by considering several linking models for across-county heterogeneity, of which the most general is the spatial Fay-Herriot model.  This model uses county-level surficial radium content (ppm), measured by the National Uranium Resource Evaluation (NURE), as an area-level predictor in a linear model.  Under this model, $\theta_1, \cdots, \theta_m$ are jointly normally distributed, with $\E{\theta_j} = \mu + \beta_1 x_j$, where $x_j$ is the measured surficial radium content for area $j$.  $\Cov{\bs{\theta}}$ is defined as in (\ref{spatiallinkingmodel}), where the proximity matrix $\bs{W}$ represents the row-standardized squared exponential distance between county centroids (measured via longitude and latitude), since no counties in Minnesota and Wisconsin are first-order neighbors of counties in Michigan or Indiana.  Explicitly,
 \begin{equation}
W_{ij} = \frac{ e^{-d_{ij}^2} }{ \left(\sum_{j \neq i} e^{-d_{ij}^2}\right)} \qquad i \neq j,
\end{equation}
where $d_{ij}$ represents the distance between the centroids of county $i$ and county $j$.  The diagonal elements of $\bs{W}$ are equal to zero.   

We also consider three simplifications of this model, corresponding to assumptions that either the regression coefficient $\beta_1 = 0$ and/or the spatial autocorrelation $\rho = 0$.  Let the matrix $\bs{X}$ be an $m \times 2$ matrix consisting of a column of ones and the column vector $(x_1, \cdots, x_m)^\top$ and let $\bs{\beta} = (\mu, \beta_1)^\top$.  Specifically, the four linking models examined are

\begin{enumerate}
\item full model: $\bs{\theta} \sim N( \bs{X}\bs{\beta}, \tau^2 [(\bs{I} - \rho \bs{W}) (\bs{I} - \rho \bs{W}^T)]^{-1})$;
\item spatial model: $\bs{\theta} \sim N( \bs{1}\mu, \tau^2 [(\bs{I} - \rho \bs{W}) (\bs{I} - \rho \bs{W}^T)]^{-1})$;
\item covariate model: $\bs{\theta} \sim N( \bs{X}\bs{\beta} ,\tau^2 \bs{I} ) $;
\item exchangeable model: $\bs{\theta} \sim N( \bs{1}\mu ,\tau^2 \bs{I} ) $.
\end{enumerate}

Unlike in the simulation study, here we treat the county-level variance parameters $\sigma^2_j$ as unknown, resulting in $t$-intervals instead of $z$-intervals.  We model the sampling variance parameters as $1/\omega^2_1,\ldots, 1/\omega^2_m \sim$ i.i.d.\ $G(a, b)$ and estimate the hyperparameters $a$ and $b$ via marginal maximum likelihood.  Details of this procedure are provided in the appendix.  Given estimates $\hat{a}$ and $\hat{b}$ based on data from other areas, we obtain prior information $1/\sigma_j^2 \sim IG(\hat{a}, n_j \hat{b})$ that is used to construct the FAB $t$-interval.  For computational convenience, we obtain prior information for $\theta_j$ separately, using plug-in estimates $\hat{\sigma}_j^2$ when estimating $\{\mu, \bs{\beta}, \tau^2, \rho\}$.  This procedure is analogous to that detailed in Section 2 and the appendix.

Because the county-level variances are unknown, we are able to construct confidence intervals with constant coverage for the 196 of the 209 counties with a sample size of at least two.  For each of these counties, we construct FAB intervals for a specific county $j$ under the four linking models specified above via the following process:
\begin{enumerate}
 \item Estimate linking model parameters using data from all counties 
   other than $j$. 
   \item Obtain prior distributions for both $\theta_j$ and 
      $\sigma^2_j$ using plug-in estimates from the fitted linking model.  This yields a normal distribution for $\theta_j$ and an inverse-gamma distribution for $\sigma_j^2$.  
\item Obtain the optimal $s$-function for county $j$ given prior information about $\theta_j$ and $\sigma_j^2$ (obtained using data not from $j$), as described in Section 2.3.
\item Construct the FAB $t$-interval $\left \{\theta : \bar{y}_j + \hat{\sigma}_j t_{\alpha(1 - s_j(\theta))} < \theta <  \bar{y}_j + \hat{\sigma}_j t_{1 - \alpha s_j(\theta)} \right \}$.
\end{enumerate}
where the quantiles correspond to a those from a $t$-distribution with $n_j - 1$ degrees of freedom.  

\begin{figure}[ht]
\centering
\includegraphics[width = 0.7\textwidth]{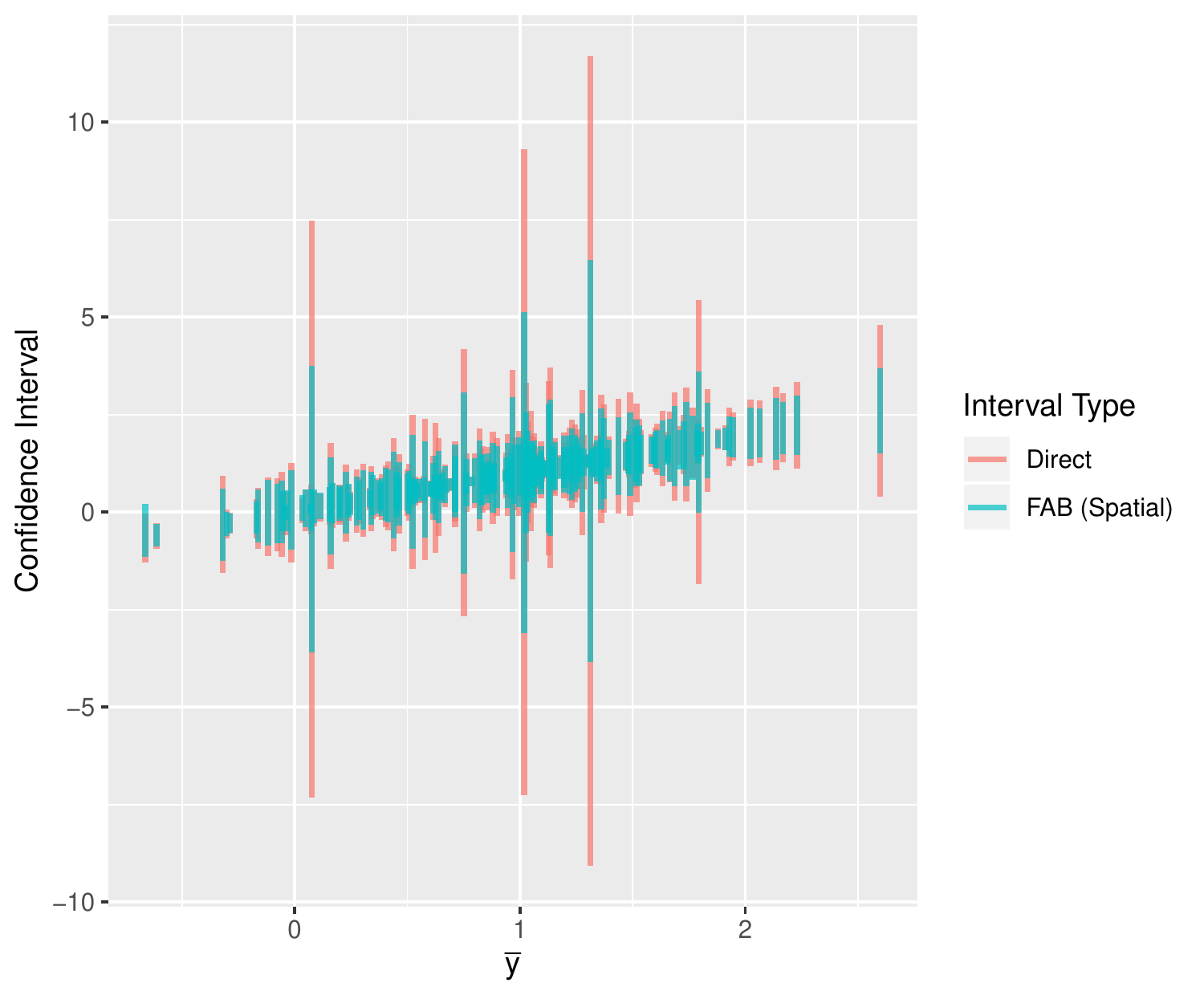}
\caption{The FAB intervals based on the spatial Fay-Herriot model are substantially narrower than the direct intervals, on average across counties in the SRRS dataset.}
\label{figure3}
\end{figure}

As visualized in Figure \ref{figure3} and depicted numerically in Table \ref{radonintervals}, FAB intervals under each of the four linking models are significantly narrower than the direct interval, representing a 23-26\% improvement in average interval width. Incorporating a spatial linking model significantly reduces interval width, and including covariate information does not appear to have much of an impact on average interval width.  Although a specific FAB interval is not guaranteed to be narrower than the corresponding direct interval, the vast majority of the FAB intervals represented improvements.  The proportion of counties with narrower FAB intervals varied from 89 to 96 percent of the counties, depending on the quality of the chosen linking model.   

\begin{table}[ht]
\centering
\begin{tabular}{|l|c|ccc|}
  \hline
Type & Linking Model  & Mean Width & Relative Width & \% Intervals Improved\\ 
  \hline
Direct &   - & 1.701 & 1.000 & -\\ 
      FAB  & Exchangeable & 1.312 & 0.771 & 89.8\%\\ 
  FAB  & Covariate & 1.312 & 0.771 & 88.8\%\\ 
  FAB  & Spatial & 1.257 & 0.739& 96.4\% \\ 
    FAB & Full & 1.256 & 0.739& 95.5\%\\ 
   \hline
\end{tabular}
\caption{Average 95\% confidence interval width, width ratio relative to the direct interval, and percentage of counties for which the FAB intervals are narrower than the direct intervals across the 196 Midwestern counties in the SRRS dataset. } 
\label{radonintervals}
\end{table}

In general, EB confidence intervals for county-specific radon levels are narrower than those constructed via the FAB procedure, although this is not always the case.  Under the full linking model, the empirical Bayes interval is narrower than the corresponding FAB interval for 128 out of the 196 counties.  The differences are most pronounced in the counties for which the combination of small sample size and high sampling variance is present.  Regardless, EB intervals lack county-specific coverage, which limits their use in making county-specific inferences.

\section{Discussion}
In the field of small area analysis, researchers typically use confidence interval procedures that either have constant coverage across areas but do not share information, or utilize shared information but lack constant coverage.  Although the empirical Bayes procedures commonly used in the literature have $1 - \alpha$ coverage on average across groups, the actual coverage rate may differ substantially for some values of $\theta_j$, calling into question the resulting area-specific inferences. The FAB procedures developed by \citeA{yuhoff16} and outlined in this article have constant $1 - \alpha$ coverage for each area regardless of what the true area-level means are, and are valid for all linear mixed models with normal sampling variances.  This class of models is very flexible, enabling researchers accommodate auxiliary covariates, as well as spatial and temporal autocorrelation.  

Importantly, although the empirical Bayes confidence interval procedure is guaranteed to have asymptotic $1-\alpha$ marginal coverage on average if and only if the linking model is true, the FAB procedure will always have $1-\alpha$ constant coverage, regardless of the chosen linking model.  This is not to say that the linking model is unimportant; a properly specified linking model can substantially reduce expected FAB interval width, as evidenced by the simulation study and empirical example.

FAB intervals are somewhat more computationally demanding to calculate than direct confidence intervals or empirical Bayes intervals since $m$ model estimations must occur to obtain confidence intervals for $m$ areas.  This can be burdensome under complex linking models, such as the spatial Fay-Herriot model,  when the number of areas is large.  Since FAB intervals will always have constant coverage, regardless of whether the linking model or the estimation procedure is correct, computational shortcuts can be taken to significantly reduce the burden, if necessary.  For example, when the number of areas is large, one possibility is to separate the areas into $k$ heterogeneous clusters and construct prior distributions for areas belonging to a given cluster based the direct estimates from other clusters.  This means that only $k$ models must be estimated, instead of $m$, resulting in computational gains.  When the number of areas is prohibitively large, we recommend simply estimating the hyperparameters of the linking model once and then using those estimates to calculate all FAB intervals.  Although this will violate the condition of independence necessary to guarantee $1-\alpha$ area-specific coverage, the influence of a single area on model estimates is likely to be small in such a context, so the FAB intervals will have very close to $1-\alpha$ coverage for all areas.     

One area of future work is to extend the FAB procedure to generalized linear mixed models by constructing FAB intervals for target quantities when responses are discrete or categorical.  This has significant applications in the small area estimation literature for applications such as disease mapping, where researchers are often interested in inferring area-level relative risks. 

The FAB procedure for constructing confidence intervals with area-specific coverage can be implemented using a variety of software packages for estimating small area estimation models, such as \texttt{sae} \cite{saepackage} or \texttt{lme4} \cite{lme4package}.  The only additional computational functionality needed is the Bayes optimal $s$-function, which we have implemented in R and made available in the \texttt{fabCI} R package on CRAN.  Replication code for the paper is provided at https://github.com/burrisk/fabci.

\clearpage

\appendix

\section{Credible interval coverage rates for the Fay Herriot model}
Under the sampling model $y_j \sim  N(\theta_j, \sigma_j^2)$ and prior $\theta_j
 \sim N(\bs x_j^\top \bs\beta, \tau^2)$, the posterior distribution

$$\theta_j \mid y_j \sim N\left(\frac{\tau^2 y_j + \sigma^2 \bs x_j^\top \bs \beta}{\sigma^2 + \tau^2}, \frac{\sigma^2\tau^2}{\sigma^2 + \tau^2}\right)$$.

Accordingly, the $1-\alpha$ symmetric credible interval $C^j_B$ can be expressed
 as

 \begin{equation*} 
C^j_B(\bs{y}) = \left \{\theta : \frac{\tau^2 y_j + \sigma^2 \bs x_j^\top \bs \beta}{\sigma^2 + \tau^2} +  \frac{\sigma\tau}{\sqrt{\sigma^2 + \tau^2}} z_{\alpha
/2} < \theta < \frac{\tau^2 y_j + \sigma^2 \bs x_j^\top \bs \beta}{\sigma^2 + \tau^2} +  \frac{\sigma\tau}{\sqrt{\sigma^2 + \tau^2}} z_{1 - \alpha/2} \right \}
\end{equation*}

For a given value of $\theta_j$, the coverage probability is

\begin{align*}
\text{Pr}(\theta_j \in C_B^j) &= \text{Pr}\left(\frac{\tau^2 y_j + \sigma^2 \bs x_j^\top \bs \beta}{\sigma^2 + \tau^2} +  \frac{\sigma\tau}{\sqrt{\sigma^2 + \tau^2}} z_{\alpha/2} < \theta_j < \frac{\tau^2 y_j + \sigma^2 \bs x_j^\top \bs \beta}{\sigma^2 + \tau^2} +  \frac{\sigma\tau}{\sqrt{\sigma^2 + \tau^2}} z_{1 - \alpha/2}\right)\\
&= \text{Pr}\left(\frac{\sigma_j (\theta_j- \bs{x}_j^\top \bs{\beta})} {\tau^2} + z_{\alpha/2} \sqrt{1 + \sigma_j^2/\tau^2} < \frac{y_j - \theta_j}{\sigma_j}\right.\\
 & \hspace{1.5cm} <  \left.\frac{\sigma_j (\theta_j- \bs{x}_j^\top \bs{\beta})} {\tau^2} + z_{1 - \alpha/2} \sqrt{1 + \sigma_j^2/\tau^2} \right)\\
&= \Phi\left(\frac{\sigma_j (\theta_j- \bs{x}_j^\top \bs{\beta})}{\tau^2} + z_{1 - \alpha/2} \sqrt{1 + \sigma_j^2/\tau^2}\right) - \Phi\left(\frac{\sigma_j (\theta_j- \bs{x}_j^\top \bs{\beta})}{ \tau^2} + z_{\alpha/2} \sqrt{1 + \sigma_j^2/\tau^2}\right),
\end{align*}

where $\Phi$ is the standard normal cumulative distribution function.

\section{ML estimation of spatial Fay-Herriot hyperparameters}
To estimate the hyperparameters $\{\bs{\beta}, \rho, \tau^2\}$ based on data from a subset of areas $S$, where $j \notin S$, we recommend using either ML or REML procedures based on the data from all areas in $S$.  We provide the details for ML estimation below, although REML estimation is straightforward, using transformed data $\bs{y}_S^* = \bs{F}^T\bs{y}_S$, where $\bs{F}$ is a $(m - 1) \times (m - p)$ matrix that is orthogonal to $\bs{X}_S$.  For more details about REML estimation for the spatial Fay-Herriot model, see \citeA{pratesisalvati08}.  Both ML and REML estimation of the spatial Fay-Herriot model are implemented in the \texttt{sae} R package and we also provide an implementation in the replication code.

Defining the marginal variance $\bs{V}_S = \bs{D}_S + \bs{G}_S$, where $\bs{G}_S = \tau^2 [(\bs{I} - \rho \bs{W}_S) (\bs{I} - \rho \bs{W}_S^T)]^{-1}$, the log-likelihood function is given by

\begin{equation*}
\ell(\bs{\beta}, \rho, \tau^2) = \text{const}- \frac{1}{2} \log{|\bs{V}_S|} - \frac{1}{2}(\bs{y}_S - \bs{X}_S\bs{\beta})^T\bs{V}_S^{-1}(\bs{y}_S - \bs{X}_S\bs{\beta}).
\end{equation*}

The MLE $\hat{\bs{\beta}}$ of $\bs{\beta}$ is of a familiar form, with

\begin{equation*}
\hat{\bs{\beta}}(\rho, \tau^2) = (\bs{X}_S^T\bs{V}_S^{-1}\bs{X}_S)^{-1}\bs{X}_S^T \bs{V}_S^{-1} \bs{y}_S.
\end{equation*}

The partial derivatives with respect to $\tau^2$ and $\rho$ are given by $s(\bs{\beta}, \tau^2, \rho)$, where

\begin{align*}
\begin{split}
s_{\tau^2}(\bs{\beta}, \tau^2, \rho) &= \frac{\partial \ell}{\partial \tau^2} \\
&= -\frac{1}{2}\text{tr}(\bs{V}_S^{-1}\bs{C}_S^{-1}) + \frac{1}{2}(\bs{y}_S - \bs{X_S}\bs{\beta})^T (\bs{V_S}^{-1}\bs{C}_S^{-1}\bs{V}_S^{-1}) (\bs{y}_S - \bs{X}_S\bs{\beta})\\
s_{\rho}(\bs{\beta}, \tau^2, \rho) &= \frac{\partial \ell}{\partial \rho} \\
&= -\frac{1}{2}\text{tr}(\tau^2 \bs{V}_S^{-1}(\bs{C}_S^{-1}[\bs{W}_S + \bs{W}_S^T - 2\rho \bs{W}_S\bs{W}_S^T]\bs{C}_S^{-1}))\\
&\hspace{4mm} + \frac{\tau^2}{2}(\bs{y}_S - \bs{X}_S\bs{\beta})^T (\bs{V}_S^{-1}(\bs{C}_S^{-1}[\bs{W}_S + \bs{W}_S^T - 2\rho \bs{W}_S\bs{W}_S^T]\bs{C}_S^{-1}) \bs{V}_S^{-1}) (\bs{y}_S - \bs{X}_S\bs{\beta}),
\end{split}
\end{align*}
where $\bs{C}_S = (\bs{I} - \rho \bs{W}_S)(\bs{I} - \rho \bs{W}_S^T)$.  We can then use these to calculate the Fisher information matrix, which is the matrix of expected second derivatives of $-\ell$.

\begin{equation*}
\mathcal{I}(\tau^2, \rho) = 
\begin{bmatrix}
\frac{1}{2} \text{tr}\left(\bs{V}_S^{-1}\bs{C}_S^{-1}\bs{V}_S^{-1}\bs{C}_S^{-1}\right) 
& \frac{1}{2} \text{tr}(\bs{V}_S^{-1}\bs{C}_S^{-1}\bs{V}_S^{-1}\bs{A}_S) \\
\frac{1}{2} \text{tr}(\bs{V}_S^{-1}\bs{C}_S^{-1}\bs{V}_S^{-1}\bs{A}_S)  
& \frac{1}{2} \text{tr}(\bs{V}_S^{-1}\bs{A}_S\bs{V}_S^{-1}\bs{A}_S) 
\end{bmatrix}
\end{equation*}
where $\bs{A}_S = \tau^2 \bs{C}_S^{-1} [\bs{W}_S + \bs{W}_S^T - 2\rho \bs{W}_S] \bs{C}_S^{-1}$.  From this, we can solve for the maximum likelihood estimates of $\tau^2$ and $\rho$ by Fisher's scoring.

\begin{equation*}
[\tau^2, \rho]^{(t+1)} = \mathcal{I}^{-1}\left([\tau^2, \rho]^{(t)}\right) \cdot s\left(\hat{\bs{\beta}}([\rho, \tau^2]^{(t)}),  [\tau^2, \rho]^{(t)}\right)
\end{equation*}
where $s$ is the $2 \times 1$ matrix of first partial derivatives with respect to $\tau^2$ and $\rho$.  Since $\tau^2$ and $\rho$ are constrained to lie in the intervals $(0, \infty)$ and $(-1, 1)$, we reduce the step size if the proposed Fisher scoring step violates one or more constraints.  The algorithm iterates until convergence. 

To obtain estimates of the subset of area means $\bs{\theta_S}$, we find their conditional means given $\bs{y}_S$, $\hat{\bs{\beta}}$, and $\hat{\bs{\psi}}$ under the sampling and linking model.  These can be expressed as 

\begin{equation*} \label{EBLUP equations}
\hat{\bs{\theta}}_{S}(\hat{\bs{\psi}}) = \bs{X}\hat{\bs{\beta}}(\hat{\bs{\psi})} + \bs{G}(\bs{\psi})\bs{V}(\bs{\psi})^{-1}(\bs{y} - \bs{X}\hat{\bs{\beta}}(\bs{\psi})).
\end{equation*}

\section{ML Estimation of sampling variance hyperparameters}
Suppose that the sampling model for the unbiased direct estimates of the area-specific sampling variances is

\begin{equation*}
\frac{(n_j - 1) \hat{\omega}_j^2}{\omega_j^2} \sim \chi^2_{n_j - 1},
\end{equation*}
where $\hat{\omega}_j^2$ is an unbiased and consistent estimate of the sampling variance $\omega_j^2$, based on a sample of $n_j$ observations.  We model the variances of log-radon levels hierarchically, under the assumption that

\begin{equation*}
1 /  \omega_j^2 \sim G(a, b), \qquad j = 1, \cdots, m 
\end{equation*}

For each area $j$, we are interested in estimating $a$ and $b$ via maximum likelihood, based on data from a subset of areas $S$, where $j \notin S$.   Then the log-likelihood is

\begin{align*}
\ell(a, b)  &= \sum_{k  \in S} \log\left(\int p(\hat{\omega}_k^2 \mid \phi_k) \text{ }p(\phi_k \mid a, b) \text{ } d\phi_k\right) \\
&= \text{const} + \sum_{k \in S} \log\int \left( \frac{b^a}{\Gamma(a)}\phi_k^{\frac{n_k - 1}{2} + a - 1} \exp\left( -\phi_k \left(\frac{n_k - 1}{2}\hat{\omega}_k^2 + b \right) \right) \text{ }d \phi_k\right) \\
&= \text{const} + |S|(a \log(b) - \log \Gamma(a)) \\
 & \hspace{1.5cm} + \sum_{k \in S} \left[\log \Gamma\left(\frac{n_k - 1}{2} + a\right) - \left(\frac{n_k - 1}{2} + a\right)\log\left(\frac{n_k - 1}{2} \hat{\omega}_k^2 + b\right) \right],
\end{align*}
where $\phi_k = 1/\omega_k^2$, $|S|$ is the cardinality of $S$ and $\Gamma(\cdot)$ is the Gamma function.  

The partial derivatives of the log-likelihood with respect to $a$ and $b$ are

\begin{align*}
\begin{split}\\
\frac{\partial \ell}{\partial a} &= |S|(\log(b) - \psi(a))  + \sum_{k \in S} \left[\psi\left(\frac{n_k - 1}{2} + a\right) - \log\left(\frac{n_k - 1}{2} \hat{\omega}_k^2 + b\right) \right] \\
 \frac{\partial \ell}{\partial b} &= \frac{|S|a}{b} - \sum_{k \in S}\left[ \frac{\frac{n_k - 1}{2} + a}{\frac{n_k - 1}{2}\hat{\omega}_k^2 + b}\right],
\end{split}
\end{align*}
where $\psi$ is the digamma function, the derivative of the log-gamma function.  In Section \ref{sec4}, we use the L-BFGS optimization algorithm with the box constraint $\{(0, \infty) \times (0, \infty\}$ to find $\hat{a}$ and $\hat{b}$, the maximum likelihood estimates of $a$ and $b$.  Due to the low dimensionality of the problem, second order information can be utilized to speed up convergence, and the second order partial derivatives are

\begin{align*}
\frac{\partial^2 \ell}{\partial^2 a} &=\sum_{k \in S}\psi^{'}\left(\frac{n_k - 1}{2} + a\right) -  |S| \psi^{'}(a) \\
\frac{\partial^2 \ell}{\partial a \partial b} &= \frac{|S|}{b} - \sum_{k \in S}  \frac{1}{\frac{n_k - 1}{2}s_k^2 + b}\\
\frac{\partial^2 \ell}{\partial^2 b} &= \sum_{k \in S}\left[ \frac{\frac{n_k - 1}{2} + a}{\left(\frac{n_k - 1}{2}s_k^2 + b\right)^2}\right] - \frac{|S|a}{b^2},
\end{align*}
where $\psi^{'}$ is the trigamma function.  An optimization algorithm that uses first-order information about $a$ and $b$ is implemented in the replication code.

\clearpage

\bibliographystyle{apacite}
\bibliography{refs} 

\end{document}